\documentclass[preprint,showpacs,prb]{revtex4}

\usepackage{amsmath} %%ADD package
\usepackage{graphicx}% Include figure files
\usepackage{dcolumn}% Align table columns on decimal point
\usepackage{bm}% bold math

\newcommand{\beq}{\begin{equation}}
\newcommand{\eeq}{\end{equation}}

\begin{document}
%%\title{Electrical readout of light polarization in a spintronics-based system}
\title{Spintronics for electrical measurement of light polarization}
\author{H. Dery}\email{hdery@ucsd.edu}
\author{\L. Cywi{\'n}ski}
\author{L. J. Sham}
\affiliation{Department of Physics, University of California San
Diego, La Jolla, California, 92093-0319}
\date{\today}
%%%%%%%%%%%%%%%%%%
\begin{abstract}
The helicity of a circularly polarized light beam may be determined
by the spin direction of photo-excited electrons in a III-V
semiconductor. We present a theoretical demonstration how the
direction of the ensuing electron spin polarization may be
determined by electrical means of two ferromagnet/semiconductor
Schottky barriers. The proposed scheme allows for time-resolved
detection of spin accumulation in small structures and may have a
device application.
\end{abstract}
\maketitle
%%%%%%%%%%%%%%%%%%
\section{Introduction} \label{sec:introduction}
The ability to measure the circular polarization of light has
greatly facilitated the progress of spintronics research in III-V
semiconductor systems \cite{Wolf_Science00} in comparison with
silicon whose conduction electrons are not optically active. The
connection between spin orientation and light polarization in
semiconductors is given by the angular momentum conservation rules.
\cite{Optical_Orientation}  The spin polarization of the electrons
in a III-V semiconductor, in the case of a current injection from a
ferromagnetic metal, has been measured either by the circular
polarization of light from a light emitting diode
\cite{Fiederling_Nature99,Hanbicki_APL02} or by scanning Kerr
spectroscopy in a lateral geometry. \cite{crooker} The spin
polarization has also been measured in the case of extraction
current into the ferromagnet by Faraday rotation, \cite{epstein} by
scanning Kerr spectroscopy, \cite{crooker} and by Hanle effect.
\cite{stephens} The determination of light polarization has, of
course, a much broader application in information technology. The
current principle of polarimetry is based on the methods of optics,
relying on, for example, photoelastic modulators and beam splitter
\cite{hipps} or the different reflection intensities of polarized
light. \cite{unlu} The question is whether one can use for
polarimetry the reversal of the roles between the ferromagnet/III-V
semiconductor and the polarized light.
The transmission from the
semiconductor to the ferromagnet is spin-dependent \cite{ciuti1} and
the spin dependence of the photocurrent passing through the junction
indicates the sense of the circular polarization of the light
exciting the electrons provided the contribution from the magnetic
circular dichroism is removed. \cite{Steinmuller_PRB05}.
% MUNEKATA ADDED:
Detection of a spin-dependent photocurrent has been recently
reported in a p-n heterojunction junction held in magnetic field
\cite{Kondo}.

\begin{figure}
       \includegraphics[height=5cm,width=7cm]{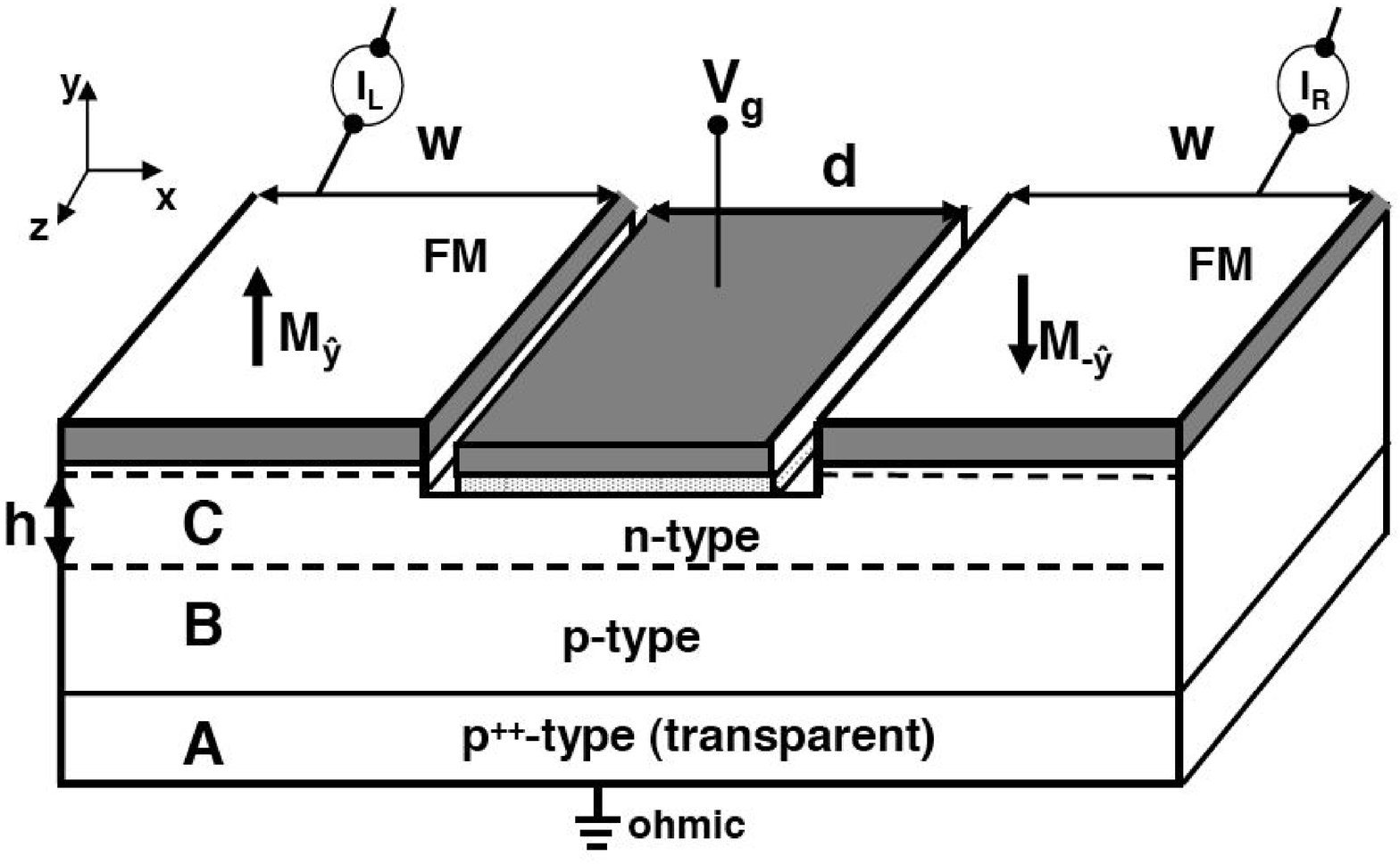}
          \caption{  \footnotesize{A scheme of the detector. The
                  conductance of the barriers is controlled by the
                  doping profile beneath the contacts. The metal gate
                  is separated form the channel by a thin insulating barrier. }}
                  \label{fig:scheme}
\end{figure}

In this paper we analyze the time dependent response of the spin
polarized electron transport after the absorption of polarized
light. We choose a spintronics system configuration which may lead
to detection of the helicity of a circularly polarized light beam as
shown in Figure~\ref{fig:scheme}. It makes a direct use of the
dependence between the magnetization of the contact and the spin
polarized electrical current across the semiconductor/ferromagnet
tunneling barrier. The beam excites spin-polarized electrons and
holes below the stack. The electrons and holes are separated by the
p-n junction. The electrons are swept upwards towards the gate and
the electrodes. The photocurrent is split through two Schottky
barriers with ferromagnets of opposite magnetizations. The different
magnitudes of the split currents depend on whether the photo-excited
electron spins are parallel or antiparallel to the magnets. The
detection scheme takes advantage of the recent advances in the
tunnel barrier fabrication either as a Schottky barrier
\cite{Hanbicki_APL02} or separated by an oxide layer \cite{parkin}
which enable efficient transmission of spin currents at room
temperature. For the potential practical applications, the
monolithic spintronics system is portable and has a much smaller
footprint than the currently available polarimeters. In addition,
the phenomenon can be used to detect the amplitude of photo-excited
spin accumulation in the semiconductor by a simple electrical
measurement. The size of the spin accumulation region may be smaller
than the spatial resolution of the scanning magneto-optical
techniques (by Faraday or Kerr effects)  limited by the wavelength
of light.

In the next section we construct a system suitable for the electron
spin response to polarized light stimulation. In section
\ref{sec:formalism} we study the transport theory of the
photo-excited electrons. We provide analytical expressions for the
steady state case whose results are presented in section
\ref{sec:results}. In section \ref{sec:time} we study the time
dependent analysis of the 2D spin diffusion equation. The results
provide a complementary physical picture for the steady state case.
We support the numerical simulations with a simplified picture from
which one can achieve a deeper understanding of the dynamics.
Conclusions are given in section \ref{sec:conclusions}.

\section{System requirements for detection of light polarization}
% Magneto-optic aspects
A polarization detection scheme would depend on the correlation
between the optical and the magnetic properties, which in turn would
require the collinearity of the light propagation axis and the
ferromagnet magnetization axes. This is simply the conservation of
angular momentum when the spin quantization axis lies along the
propagation direction of the polarized light. In terms of the axes
labeled in Fig.~\ref{fig:scheme}, the collinear direction is along
the $y$-axis in which pinning of the magnetization out of plane may
be realized by stacking alternate ferromagnetic and
antiferromagnetic layers on top of each
contact.\cite{garcia_out_of_plane}
% ADDED:
The light can come from below, through the transparent substrate, in
which case the thickness of the absorbing layer should not exceed
the absorption depth. Alternatively, the light can come from above,
and depending on the thicknesses of the contacts and the gate (see
Fig.~\ref{fig:scheme}), the photocarriers can be generated
everywhere in the semiconductors channel, or only under the gate.

% chosen semiconductor material
The choice of the semiconductor materials is governed by the
wavelength of the excitation light. In bulk semiconductors, the
upper and lower bounds of the wavelength are set, respectively, by
the threshold of band gap transitions and by the onset of
transitions between the split-off valence band and the conduction
band. For circularly polarized beam the degeneracy of the heavy and
the light hole bands in bulk III-IV compounds leads to our
assumption of optical excitation of both types of holes. In order to
quantify the spin polarization the beam intensity is decomposed into
two generating terms of photo-excited electrons $I_+$ and $I_-$
which correspond, respectively, to electrons with spin up and spin
down. The resulting spin polarization is
$\rho$$\equiv$$(I_+$$-$$I_-$$)$/$(I_+$$+$$I_-$$)$$=$$\pm$$1/4$ where
the sign depends on the light helicity. \cite{Optical_Orientation}

% gate functionality
%THE NEXT PARAGRAPH IS FOR MAGNETISM PEOPLE WHO MIGHT NOT UNDERSTAND
%THE REASON FOR PUTTING A GATE
The gate is essential to the stability of the system. It screens the
in-plane electric field due to excess electrons in the channel by
bringing ``mirror'' charges to the metal surface adjacent to the
insulator. The screening is effective for a high aspect ratio
between the channel length and the thickness of the insulating
barrier separating it from the gate. Consequently, the system is
electrostatically stable as in MOSFET devices and the transport
along the axis connecting the ferromagnets may be assumed to be
purely diffusive. The gate may have an additional role if the p-n
junction of Fig.~\ref{fig:scheme} is replaced by an alternate design
with a unipolar doped structure. In this case the gate should be
biased so that a charge accumulation (inversion) layer is formed if
an n-doped (p-doped) semiconductor is used. The gate electric field
in the semiconductor replaces the built-in field of the p-n junction
in the function of sweeping the electrons into the conduction
channel. Note that the electron gas in the conduction channel is 3D
in the quantum character as we take the effective conduction channel
thickness to exceed the electron de Broglie wavelength.

The scale of the system geometry is set by the spin diffusion
length, about 1~$\mu$m in GaAs at room temperature, which limits the
travel distance of the spin-polarized electrons in the system. For
high detection efficiency, the thickness of the conduction channel
should be much less than the spin diffusion length. The width of the
gate, which attracts the excited electrons to the conduction
channel, must also be less than the diffusion length. Below we find
the optimal width of each ferromagnetic contact to be less than the
spin diffusion length. There is a corresponding important dependence
of the efficiency of the lateral semiconductor spin valve on the
width of the contacts in the planar geometry. \cite{Dery_contact} In
order to limit the spins to the active area, the pillar structure is
designed to restrict the electrons under the ferromagnetic contacts
or between them.

\section{Electron transport} \label{sec:formalism}

The polarized beam creates non-equilibrium spin-dependent components
of the density $\delta n_s$, $s$$=$$\pm$ for spin up or spin down.
The resultant net spin density, $\delta n_{+}$$-$$\delta n_{-}$,
gives rise to spin accumulation in the channel. To simplify the
following numerical procedures and in order to extract analytical
expressions when possible we consider weak excitations. This means
that the deviations from equilibrium, $\delta n_s$, are one or two
orders of magnitude smaller than the free carrier concentration in
the paramagnetic channel at equilibrium ($n_0$). In this regime the
electrochemical potential $\mu_s$, may be taken to be linear in
$\delta n_s$ in addition to the electric field term. However the
screening action of the gate renders the contribution of this field
negligible for transport inside the conduction channel (between the
ferromagnets). After excitation, diffusion currents start flowing
into the tunneling contacts, equivalent to currents under low
forward bias to the Schottky barriers but driven by the spin density
gradient, thus,
 \begin{equation}
\mathbf{j}_s = \frac{\sigma_{s}}{e} \nabla \mu_s,
\end{equation}
where $-e$ is the electron charge and  the conductivity in each spin
channel is half of the total conductivity of the semiconductor,
$\sigma_{sc}$. The density profiles are related to the currents by
the spin dependent continuity equations,
\begin{equation}
\frac{\partial \delta n_s}{\partial t} = \frac{1}{e}\nabla
\mathbf{j}_s - \frac{\delta n_s - \delta n_{-s}}{2\tau_{sp}}  + I_s
\,\, ,\label{eq:continuity}
\end{equation}
where $2\tau_{sp}$ is the average flip time between two spin states
and $I_s$ is the spin dependent optical generation rate derived from
light intensity and absorption coefficient. Combination of the two
equations leads to the time-dependent diffusion equation,
\begin{eqnarray}
\frac{1}{D}\frac{\partial \mu_{s}(x,y)}{\partial t} = \nabla^2
\mu_{s}(x,y) - \frac{\mu_s(x,y) - \mu_{-s}(x,y)}{2L_{sc}^2} +
\frac{2k_BT}{Dn_0}I_s \,\, , \label{eq:cont2}
\end{eqnarray}
where $D$ is the semiconductor diffusion constant for the single
spin component and $L_{sc}$$=$$\sqrt{D \tau_{sp}}$ is the
semiconductor spin diffusion length ($L_{sc}$). The temperature
factor comes from our application to the specific case of a
non-degenerate semiconductor channel. For the degenerate electrons,
it can be easily replaced by the inverse of the derivative of the
chemical potential (i.e., the compressibility of the electron gas).

The boundary condition for the normal component of the spin current
across the ferromagnet/semiconductor interface is given by,
\begin{eqnarray}
\sigma_{sc} \Big(\widehat{n}\cdot\nabla\mu_{s}\Big) =   G_s (-eV -
\mu^i_s), \label{eq:boundaries}
\end{eqnarray}
where $\widehat{n}$ denotes the outward interface normal from the
semiconductor and $G_s$ is the spin dependent barrier conductance
per unit area ($\Omega^{-1}$cm$^{-2}$). $V$ is the bias voltage
applied to the ferromagnetic contact above this part of the channel.
We have replaced the exact description of the electrochemical
potential in the ferromagnet with the bias level, justified by the
vastly different conductivities of the ferromagnetic metal and the
semiconductor ($\sigma^{{fm}}_{s}$$\gg$$\sigma_{sc}$). In the middle
part of the channel the leakage current into the gate is negligible
for an insulator layer over 5~nm and so the boundary conditions at
the interface are $j^{y}_s$$=$$0$.
%%% I have moved the sentence about the interface scattering form
%the end of the previous papragraph to here, as it seems a more logical place:
We neglect interfacial spin scattering since it is less important
than the spin selectivity of the tunneling transmission and, in any
case, can be incorporated phenomenologically by the spin-dependence
of the barrier conductance \cite{Dery_contact}.
%%%

In the conduction channel, the effective spin generation rate,
$I_s$, has a small fraction from direct absorption of light and a
major part from photo-excited electrons driven in from the diode
region by a strong electric field along the $y$ axis (by the
definition of the axes in  Fig.~\ref{fig:scheme}). In the depleted
region, the electric field ($|$$e$$E_y$$|$$\gg$$k_BT/$$L_{sc}$)
enhances the downstream spin diffusion length by orders of magnitude.
\cite{Yu_Flatte_long_PRB02,Smith_book} Thus, spin flip processes
play negligible role in this region and the carriers flushed across
the $y$$=$$h$ plane retain their spin orientation. The effective
optical generation rate in the channel $I_{s}$ is thus up-scaled by
a factor of $(h$$+$$H)$$/$$h$, where $H$ is the depletion region
width.

The diffusion spin current in the channel is reduced to one
dimension. \cite{Dery_contact} With the axes defined in
Fig.~\ref{fig:scheme}, there is no $z$-dependence. The two
dimensional flow ($x$ and $y$ dependent) in the channel is reduced
to one dimensional along it driven by the vertical ($y$) average of
$\mu_{s}$
\begin{eqnarray}
 \xi_{s}(x)=\frac{1}{h}\int_0^h dy \, \mu_{s}(x,y)\,, \label{eq:avg_mu}
\end{eqnarray}
where $h$ is the conduction channel thickness, bounded by the gate
and the onset of the depletion region for a $p$-$n$ junction or
equivalent for the other mentioned cases. The key two dimensional
character retained is the $x$ dependence of the current along the
injection contact, a property which is lost in the usual collinear
contact/channel/contact one-dimensional model. The present
approximation is valid when $h$$\cdot$$G_s$$\ll$$\sigma_{sc}$,   a
condition fulfilled in the system under discussion. For steady state
and for homogeneous excitation we derive an effective 1D equation
governing the lateral spin transport. Integrating out the $y$
dependence in Eq.~(\ref{eq:cont2}) and using
Eq.~(\ref{eq:boundaries}) yields
\begin{eqnarray}
\frac{\partial \xi_{s}(x)}{\partial x^2} &=& \frac{\xi_{s}(x) -
\xi_{-s}(x)}{2L_{sc}^2} + \frac{2G_s}{\sigma_{sc} h} ( eV +
\xi_{s}(x)) - \frac{2k_BT}{Dn_0}I_s \,, \label{eq:diffusion_avg}
\end{eqnarray}
where $V$ is the voltage bias on the contact. The term containing
$G_s$ is omitted for the middle section beneath the gate. We present
now the spin dependent solutions of Eq.~(\ref{eq:diffusion_avg})
under the left and right contacts, occupying, respectively, $x
\in[0,w]$ and $x \in$$[$$w$$+$$d$$,$$2$$w$$+$$d$$]$. $w$ is the
width of the contacts, and $d$ is the width of the gate (distance
between the contacts), see Fig.~\ref{fig:scheme}. For a compact form
of the solutions, we define inverse diffusion lengths,
\begin{eqnarray}
\lambda^2_{(s,c)} &=& [\alpha+ 1 \pm  \sqrt{1+
\beta^2}]/(2L_{sc}^2),
\end{eqnarray}
and the following dimensionless parameters,
\begin{eqnarray}
\lambda &=& \cot{\bigg[\ \frac{1}{2}\text{tan}^{-1}{\beta} \Bigg]},
\nonumber \\ ( \alpha,\beta) &=& 2L_{sc}^2(G_+\pm
G_-)/(\sigma_{sc}h_{_{sc}}),
 \label{eq:cont_def}
\end{eqnarray}
where the first of each pair of symbols $(s,c)$ or $(\alpha,\beta)$
takes the upper sign. In the presence of weak excitation intensities
the shape of the tunneling barriers is nearly unaffected. Thus, the
antiparallel alignment of the magnetization results with
$\alpha_L$=$\alpha_R$ and $\beta_L$=$-\beta_R$ where L and R denote,
respectively, the left and right tunneling barrier. This symmetry
holds if the applied bias is much smaller than the Schottky barrier
height (e.g. $|V_{L}$$-$$V_{R}|$$<$$0.1V$ for Fe/GaAs structures).
Also the generation rates are rewritten as
\begin{equation}
I_{T} = I_{+} + I_{-} \qquad \qquad I_{D} = I_{+} - I_{-}.
\label{eq:cont_def2}
\end{equation}
The \textit{steady state} solution beneath the contacts for
spatially uniform excitation is given by
\begin{eqnarray}
\xi_{\pm}\!\!\!\!\!&&\!\!\!(\!x\!) \!=\! (1 \! \pm \! \lambda) \!
\Big[\!Ae^{\lambda_{s}x}\!\!+\!\!Be^{-\lambda_{s}x}\!\Big] \!+\!
(\lambda \! \mp \! 1)\! \Big[\!
Ce^{\lambda_cx}\!\!+\!\!Fe^{-\lambda_cx} \! \Big] \nonumber \\
&& -eV_{L(R)} + \frac{2k_BTL^2_{sc}}{Dn_0} \cdot \frac{ ( \alpha \mp
\beta )(I_T \pm I_D ) + 2I_T}{ \alpha^2 - \beta^2 + 2\alpha }\,.
\label{eq:cont_final}
\end{eqnarray}
The first line denotes the homogenous solution in which the
transport is characterized by two length scales ($\lambda^{-1}_{s}$
and $\lambda^{-1}_{c}$). Consider the case that spin polarization is
robust so that $\alpha$ and $\beta$ are comparable. If $\alpha \ll
1$, then $\lambda_c \ll 1/L_{sc}$ and  $\lambda_s \sim 1/L_{sc}$.
The $s$-mode is limited by the spin diffusion constant and it
corresponds to spin accumulation ($\lambda$$\gg$$1$ in this case).
If $\alpha \gg 1$, then both eigenvalues are nearly independent of
$L_{sc}$, and neither of the eigenvectors is a pure spin mode
$\lambda$$\simeq$$1$: the inhomogeneity of extraction dominates the
spatial dependence. The second line denotes the inhomogeneous
solution and is null for zero bias and when excitation is allowed
only beneath the gate. In the latter region, the steady state
solution is:
\begin{eqnarray}
\xi_{\pm}(\!x\!) \!&=&\! A_c \!+\! B_cx \! \pm \! \Big( C_c
e^{\frac{x}{L_{sc}}} \! + \! F_c e^{-\frac{x}{L_{sc}}} \Big)
\nonumber \\ && - \frac{k_BT}{Dn_0}\Big(\frac{I_T}{2}x^2 \mp I_D
L_{sc}^2\Big) \,.\label{eq:steady_state_channel}
\end{eqnarray}
The term quadratic in $x$ is due to the locally  uncompensated
charges in the channel, which are neutralized by the gate.
\cite{Smith_book} The coefficients of the homogeneous solutions are
determined by joining $\xi_{s}$ and its first derivative at the
boundaries between the sections of the channel ($x$$=$$w$ and
$x$$=$$w$$+$$d$). These conditions correspond to the continuity of
densities and currents. In addition, the derivatives vanish at the
outer boundaries ($x$$=$$0$ and $x$$=$$2w$$+$$d$). The total current
amplitudes in the ferromagnets are:
\begin{eqnarray}
I_L &=& Z   \sum_{s=\pm} \int_0^w \frac{G^L_s}{e} \Big[\xi_{s}(x)
+eV_L \Big] dx \nonumber \\ I_R &=& Z \sum_{s=\pm} \int_{w+d}^{2w+d}
\frac{G^R_s}{e} \Big[\xi_{s}(x) +eV_R \Big] dx \label{eq:IL_IR}
\end{eqnarray}
where $Z$ denotes the contact length along the $z$ direction.

\section{Results and discussion in steady state} \label{sec:results}
The key discriminant for the light polarization direction is the
current asymmetry (CA) coefficient $=|I_L/I_R-1|$, where we assume
that the currents have been balanced by a small voltage adjustment
when the light is unpolarized.  We have performed our calculations
using the parameters of a GaAs/Fe system at room temperature. The
equilibrium electron concentration in the n-type GaAs channel is
taken to be $n_0$$=$$10^{15}$ cm$^{-3}$ and the effective channel
thickness $h$$=$$100$ nm. The barrier conductances are
$G_{+}$$=$$2$$G_{-}$=$500$ $\Omega^{-1}$cm$^{-2}$ for one contact,
and the roles of $+$ and $-$ are switched for the other one. The
$G_{+}/G_{-}$ ratio and the overall order of magnitude of $G_{s}$
agree with the polarization and I-V measurements in spin LEDs.
\cite{Hanbicki_APL03} In the channel, the semiconductor diffusion
coefficient $D$ is 180~cm$^2$/sec and the mobility $\nu$$=$$7000$
cm$^2$/V-sec. The semiconductor spin relaxation time is
$\tau_{sp}$$=$$80$ ps. \cite{Optical_Orientation} Accordingly, the
dimensionless quantity $\alpha$=$2$. We define the barrier's finesse
$F$$\equiv$$\beta/\alpha$, $-1$$\leq$$F$$\leq$$1$, describing the
spin selectivity of the barriers ($|F|$$=$$1$ for the perfect spin
injection from a half-metallic ferromagnet). With the reasonable
parameters above, $|F|$$=$$1/3$.

\begin{figure}
\includegraphics[height=8.5cm,width=8.5cm]{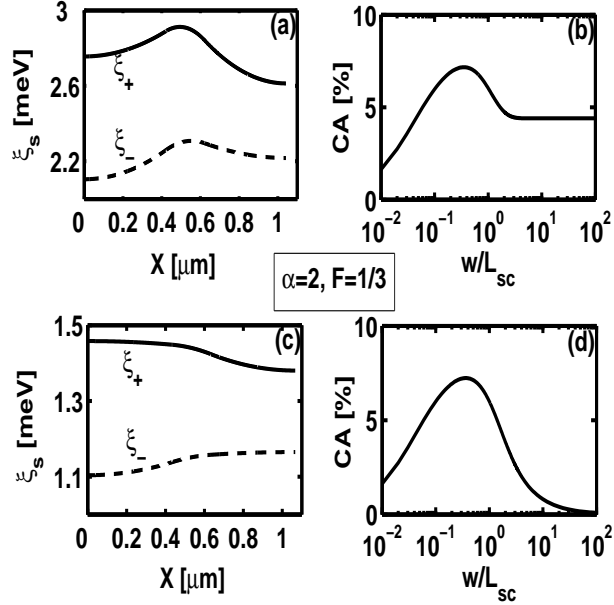}
\caption{ \footnotesize{ (a) and (c) show, respectively, the steady
state profiles of the electrochemical potentials for excitations
restricted only beneath the gate and without such restriction. The
spin-up (down) is parallel (antiparallel) to the majority spin axis
in the the right ferromagnet. The right current ($I_R$) is
\protect{$7\%$} higher than the left current ($I_L$). (b) and (d)
show the respective current asymmetry (CA) coefficient as a function
of the ferromagnetic contact width $w$. For all cases, the
temperature is 300K, \protect{$V_{fm}^L$=$V_{fm}^R$} and the same
weak intensity level is used. The sub-gate region is centered and
extends over 200~nm while the rest represents the sub-contact
regions. . }} \label{fig:mu_and_CAvsW}
\end{figure}

Fig.~\ref{fig:mu_and_CAvsW}a shows the electrochemical potential
profiles in the semiconductor channel for excitation level at which
the nonequilibrium density is $\delta n_+$$+$$\delta
n_{-}$$\simeq$$0.04n_0$.
%%% LIGHT POWER ADDED:
This corresponds to light with $\hbar \omega$$=$$1.5$ eV (band gap
of GaAs) and power of about 1 W/cm$^{2}$, when the light is absorbed
in a layer of 1 $\mu$m thickness and all the photoelectrons are then
drawn into the 100 nm thick n-type channel.
%%%
The excitation is restricted only to the sub-gate region so that the
inhomogeneous term in the second line of Eq.~(\ref{eq:cont_final})
vanishes. The separation of the ferromagnets is $d$$=$$200$ nm which
is well within the present planar lithography resolution. The
contact widths are $3L_{sc}/8$$=$$450$ nm. Using these parameters we
obtain a $\sim$$7$$\%$ difference between the left and right
ferromagnet currents. We note that the slope of spin-up (spin-down)
electrochemical potential is steeper toward the right (left)
ferromagnet due to more efficient electrons extraction of this spin.
Flipping the helicity results in a mirror image of the spatial
profile and in switching roles $\xi_+$$\leftrightarrow$$\xi_{-}$.
The larger current is extracted from the side with the larger spin
depletion. Fig.~\ref{fig:mu_and_CAvsW}b shows the dependence of CA
on the ferromagnetic contact width $w$ with an optimal value
relative to the spin diffusion length. Figs.~\ref{fig:mu_and_CAvsW}c
and ~\ref{fig:mu_and_CAvsW}d show the respective results where the
excitation is allowed beneath the contacts and the gate together.
The existence of a peak in the CA may be understood by the behavior
in two extreme cases. For contacts whose width exceeds the spin
diffusion length the behavior is different for both excitation
cases. If the excitation is restricted only beneath the gate region
than the asymmetry reaches a finite asymptotic value as the spin
information is already lost when diffuses beyond this width scale.
For non-restricted excitation the fraction which contributes to the
asymmetry vanishes if $w$$\gg$$L_{sc}$ and electrons would tunnel
through the barrier under which they were generated. The other
extreme of small contacts ($w$$\ll$$L_{sc}$) results with similar CA
dependence on the contact width as most of the electrons are excited
beneath the gate. We mention that although the principle of
operation of our proposal is straightforward, it is different than
the existing electrical measurement scheme with a single
ferromagnetic contact on top of the semiconductor layer
\cite{Hirohata_PRB01,Isakovic_JAP02}. If the contacts are separated
by less than the spin diffusion length than the spin accumulation
profile in the channel ''senses`` the antiparallel contacts.
Consequently, the difference in the relative current magnitude from
each terminal is sharper compared to the difference in photocurrent
from a single contact scheme.

\begin{figure}[h] %[!p]
\includegraphics[height=8cm,width=10cm]{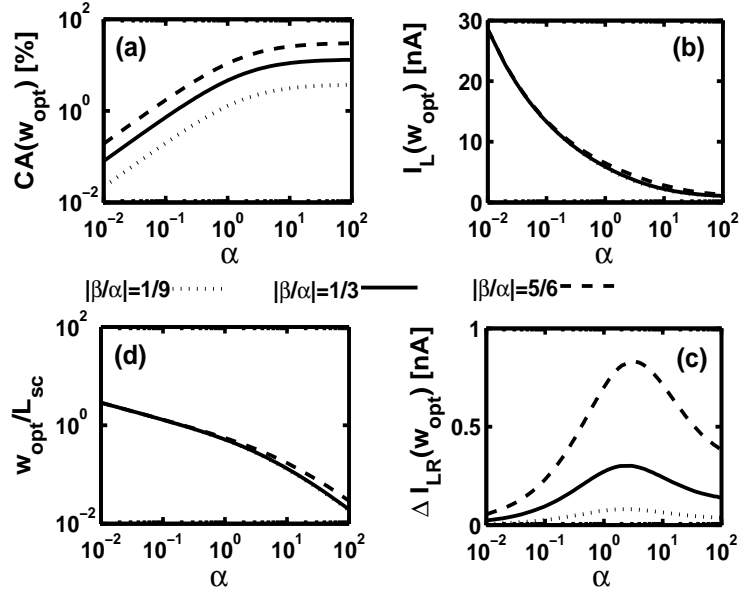}
\caption{  \footnotesize{(a) current asymmetry versus
\protect{$\alpha$} for three cases of spin selectivity. (b) and (c)
are, respectively, the current from the left contact and the
magnitude of the current difference between the left and right
contacts. All calculations are done for a structure with an optimal
contact widths shown in (d). The spin diffusion length is
\protect{$L_{sc}=1.2~\mu$m} and the separation between the
ferromagnetic contacts is \protect{$d= 0.2~\mu$m}.\label{fig:CA} }}
\end{figure}

The effect of the barrier's conductance ($G_{s}$) on CA may be
studied directly by varying the doping profile beneath them.
Fig.~\ref{fig:CA}a shows the value of CA for optimal value of the
contact width as a function of dimensionless parameter $\alpha$
defined in Eq.~(\ref{eq:cont_def}) for three different values of
barrier finesse $|F|$. In all of the following, the excitation is
allowed in all regions. The  fixed parameters are the spin diffusion
length $L_{sc}=1.2~\mu$m and the ferromagnets' separation
$d=L_{sc}/6$. Curves for different values of $\rho$ and $F$ show
that the CA is proportional to $\rho F$ and is independent of the
excitation level in the linear regime ($|\delta n_\pm|<<n_0$). The
linear dependence on the finesse comes from the difference between
two currents being linear in the spin selectivity. We recall that in
lateral spin valves electrons traverse two barriers leading to
quadratic dependence of magneto-resistive effect on F. In the
detector scheme the role of one ferromagnetic contact (injector) is
replaced by the photexcitation process ($\rho$), and when properly
designed the carriers may ``select'' their preferable extracting
terminal.

The optimal contact widths $w_{opt}$ for which these CA values were
obtained are plotted in Fig.~\ref{fig:CA}d.
%From the lowering of $w_{opt}$ as  $\alpha$ increases, we can infer
%that CA would decrease for a fixed $\alpha$ as $w$ increases from its optimalvalue.
The lowering of $w_{opt}$  with increasing $\alpha$  is understood
as follows. For highly conductive contacts, the inhomogeneity of
extraction dominates the spatial dependence of spin densities. This
means that electrons coming from the gate region will immediately
leave the channel when reaching the contact as it would be the path
with minimal resistance. Moreover, due to the finite spin dependent
conductance and the overall high conductance, photo-excited
electrons which are being created beneath the far edge of the
contact will rather leave from the same edge leading to further
reduction in the asymmetry. This is in contrast to the case of low
conductances where the current extraction profile is homogeneous
beneath the gates. Finally, we note that $w_{opt}$ is weakly
dependent on the finesse of the barriers. This observation
simplifies the procedure for choosing the contact width as $\alpha$
can be easily obtained from the I-V curve of the junction without
need of the knowledge of $\beta$, which is harder to measure.

Fig.~\ref{fig:CA}b and Fig.~\ref{fig:CA}c show, respectively, the
current from the left contact and the magnitude of the current
difference between the left and right contacts for excitation with
left circular polarization. The semiconductor parameters and the
light intensity are the same as before
(Fig.~\ref{fig:mu_and_CAvsW}). The contact length along the z axis
of Fig.~\ref{fig:scheme} is 1~$\mu$m. Although we are using the same
excitation level the current is reduced with $\alpha$ as the optimal
contact width becomes narrower. For the case of $|F|$=5/6 and
$\alpha \geq 10$ the CA exceeds 30\%.

\section{Time dependent analysis} \label{sec:time}

\begin{figure}[h]
\includegraphics[height=5cm,width=7.5cm]{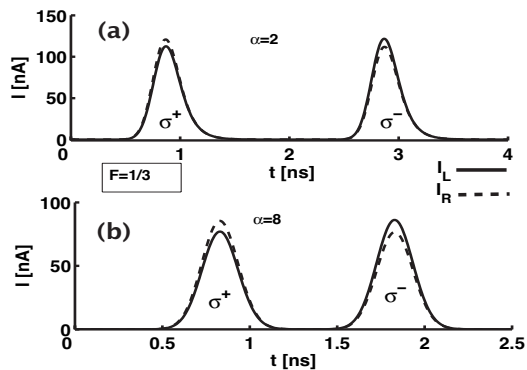}
\caption{ \footnotesize{Time-resolved current response to light
pulses alternating in polarization. The solid (dashed) line is from
the left (right) magnetic contact. (a) Low conductance barriers with the light pulse rate is 0.5~GHz and the
two pulses centered at 0.8~ns and 2.8~ns.  (b) Barriers of higher conductance with the
repetition rate is 1~GHz with two pulses centered at 0.8~ns and
1.8~ns.}}\label{fig:pulse}
\end{figure}

The circular polarization of a pulse of light may also be determined
by this spintronics system. We start by a numerical simulation of
the time-dependent diffusion equation (\ref{eq:cont2}). The initial
condition corresponds to a quiescent medium: $\delta
n_s$$(x,y,t=0)$$=$$0$. Fig.~\ref{fig:pulse} shows the currents
through the two ferromagnetic contacts as a result of excitation by
two consecutive Guassian-shaped pulses of opposite polarization. The
width of the pulses is $100$ ps. Figure \ref{fig:pulse}a shows a
0.5~Ghz repetition rate, with all the parameters as in
Fig.~\ref{fig:mu_and_CAvsW}, and using the average light power of 5
W$/$cm$^{2}$ (calculated using the same assumptions as in the
steady-state case). In Fig.~\ref{fig:pulse}b  we use barrier
conductances four times larger and 1 GHz repetition rate with
average light power of 10 W$/$cm$^{2}$. In both cases, the peak
power is 40 W$/$cm$^{2}$. The contact widths have been optimized in
both cases according to the steady state analysis above.
Accordingly, in the higher conductance barriers case (lower panel)
the contact widths are equal to $L_{sc}$$/$$6$. The contact length
along the z axis in Fig.~\ref{fig:scheme} is 1~$\mu$m. The rise of
the current signals follows the light pulses (not shown), although
there is $\sim$$65$ ps ($\sim$$30$ ps) lag between the peak of the
light pulse and the peak of the current in Figure
\ref{fig:pulse}a(b).
%%% FLUSH-OUT TIME ADDED:
This lag is caused by a finite time which is needed for electrons to
leave the channel by tunneling through the barriers.

A simple estimate of a time associated with the flush-out of the
photoexcited carrier density  is obtained in the following way.
Firstly, we set up equations for time-dependence of spin densities
after an instantaneous excitation by neglecting the diffusion inside
the channel and assuming a spatially uniform distribution.
Integrating out the $x$ and $y$ coordinates in the continuity
equation (Eq.~(\ref{eq:continuity})) yields,
\begin{equation}
\frac{\partial \delta n_{s}}{\partial t} = \frac{1}{e}
\frac{w}{(2w+d)h} (j_{L}+j_{R}) - \frac{ \delta n_{s} - \delta
n_{-s}}{2 \tau_{sp}} \,\, ,
\end{equation}
where $j_{R}$ and $j_{R}$ are the averaged current densities at the
contact interfaces. In order to express the currents in terms of the
non-equilibrium densities, we make use of Eq.~(\ref{eq:boundaries})
and of the relation between the chemical potential and the
non-equilibrium density in the linear regime. % $\mu_{s}$$=$$ 2kT
%\delta n_{s}/n_{0}$, with total equilibrium carrier density $n_{0}$.
We arrive at the following equations for the total photoexcited
density $\delta n$=$\delta n_{+}+\delta n_{-}$ and for the
photoexcited density polarization $\Delta n$$=$$\delta n_{+}-\delta
n_{-}$:
\begin{eqnarray}
\frac{\partial}{\partial t} \delta n =  - \frac{\delta n}{\tau_{f}}
\,\,. \qquad \frac{\partial}{\partial t} \Delta n  =  - (
\frac{1}{\tau_{f}} + \frac{1}{\tau_{sp}}) \Delta n\,,
\end{eqnarray}
where the carriers ``flush-out'' time $\tau_{f}$ is given by
\begin{equation}
\tau_{f} = \frac{(2w+d)h}{w} \frac{\sigma}{2D(G_{+}+G_{-})}=
\frac{2w+d}{w} \frac{\tau_{sp}}{\alpha} \,\, .
\end{equation}
The total current out of the system is proportional to $\delta n$,
so it decays exponentially with time-constant $\tau_{f}$. This time
constant is spin independent and relates only to the total
resistances of the FM/semiconductor barriers and of the
semiconductor layer ($\alpha$ $\propto$ $\tau_{sp}$). The difference
$I_{L}-I_{R}$ is proportional to the spin accumulation $\Delta n$,
and decays on a time-scale of $\tau_{LR}^{-1} $$=$$
\tau_{sp}^{-1}+\tau_{f}^{-1}$. These results agree very well with
the numerical calculations using the full time-dependent diffusion
equation. We can see that the time-scale after which an excited
system returns to its equilibrium state is given by $\tau_{f}$,
which limits the repetition rate of light pulses. It is also
favorable to have $\tau_{f}$$<$$\tau_{sp}$, so that the
photoelectrons leave the channel before losing their spin
polarization; the time-scale on which the CA effect disappears is
bounded from above by the spin relaxation time $\tau_{sp}$. However,
too short $\tau_{f}$ is also undesirable. The explanation of this
leads to an alternative understanding of the optimal contact size
discussed for the steady state case.

In order to analyze the effect of short $\tau_{f}$, we have to relax
the simplifying approximation of $\delta n_{s}$ uniformity and
reintroduce the diffusion processes. A typical time for the density
perturbation to propagate through distance $l$ is
$\tau_{\text{diff}}$$\sim$$l^{2}/D$. If the carriers tunnel into
each of the contacts faster than they diffuse between them, both
contacts do not ``sense'' each other. The time-integrated CA will
vanish in such a case, as electrons leave the channel through the
nearer contact, and the average $I_L$ and $I_R$ currents will be the
same. The requirement for $\tau_{f}$ to be smaller than $\tau_{sp}$
but larger than $\tau_{\text{diff}}$ leads to inequalities:
\begin{equation}
\frac{l^2}{L_{sc}^2}\alpha < \frac{2w+d}{w} < \alpha  \,\, ,
\end{equation}
where $l=w+d$ is a typical distance on which an electron has to
diffuse to get from under one contact to another. From the above
inequalities, we can qualitatively recover the results of
Fig.~\ref{fig:CA}a and Fig.~\ref{fig:CA}d. For $\alpha < 2$, spin
relaxation is faster than the flush-out time and the right hand side
of the inequality is violated. Consequently, further reduction of
$\alpha$ results in weaker CA effect in agreement with the steady
state behavior shown in Fig.~\ref{fig:CA}a. In the limit of large
$\alpha$, it is possible to satisfy the left hand side of the
inequality by shrinking the contact width compared with the gate
width: $w<d$. The diffusion process is faster than the flushing time
if $w$ is smaller than $L_{sc}^{2}/(d\alpha)$. The spin flip
processes are of no importance on these time scales as long as
$L_{sc}>d$ so that the right part of the inequality is fulfilled.
The CA is maximal in this regime, and $w_{opt}$ $\propto$
$1/\alpha$, as one can see in Fig.~\ref{fig:CA}d. Satisfying both
conditions of the inequality results in high CA effect which in
steady state corresponds to the plateau of Fig.~\ref{fig:CA}a. The
time domain analysis clarifies the relatively weak dependence on the
finesse. This is seen from the flush-out time, the diffusion time,
and the spin relaxation time which involve only $\alpha$, $L_{sc}$,
and the length scales of the structure in hand.
%%%

\begin{figure}[h]
\includegraphics[height=6cm,width=7.5cm]{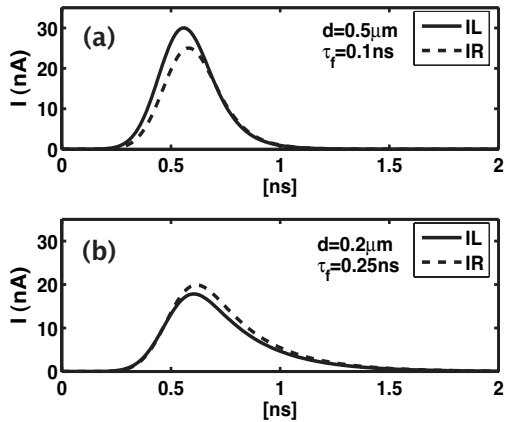}
\caption{ \footnotesize{Time-resolved current responses to
excitation of a 0.1~ns wide gaussian pulse centered at 0.5~ns. The
light is right circularly polarized and the photoexcitation occurs
under the left contact.}}\label{fig:pulse2}
\end{figure}

To study the competition between the diffusion and flush-out
processes we consider the excitation only under the left contact.
%LET'S NOT BE THAT SPECIFIC: BELOW WE RELY ON THE MODEL OF
%SPIN SELECTIVITY THAT WE USE, BUT GENERALLY IT COULD BE THE OTHER
%WAY AROUND. AND MOST OF THE PEOPLE GET CONFUSED BY SPIN DIRECTIONS
%VS MAJORITY/MINORITY AND THE DIRECTION OF MAGNETIZATION.
%The circular polarization creates photoelectrons with net spin in
%parallel to the right contact majority spin.
% REPLACED WITH:
The circular polarization of light is such that for uniform
excitation in the channel the current from the right contact should
be larger. In case (a), we simulate a structure whose total barrier
conductance is 1000 $\Omega^{-1}$cm$^{-2}$ and whose gate width is
500~nm. In case (b), the respective values are 300
$\Omega^{-1}$cm$^{-2}$ and 200~nm. In both cases, the contact widths
are 400~nm and the spin relaxation time is 8~ns (possible at lower
temperature in GaAs) so that electrons leave the channel without
losing their spin information. Other parameters are the same as
before. The two cases are presented in Fig.~\ref{fig:pulse2}a and
\ref{fig:pulse2}b, where the time-resolved current signal is shown
for excitation by a 0.1~ns wide gaussian pulse centered at 0.5~ns.
We see that in case (a) the diffusion process is not capable of
counteracting the spatial inhomogeneity in excitation and the
current from the left terminal is stronger due to the fast flush-out
time regardless of the unfavorable polarization. On the other hand,
in case (b) the shorter channel and lower barrier conductance enable
the diffusion process to fix the spatial inhomogeneity in
excitation, and the right terminal manages to extract more carriers
as one would expect for homogenous excitation.

In the non-biased system the noise is governed by thermal
fluctuations, as the shot noise is irrelevant at predicted current
levels. The pulses of 100~ps amount to a bandwidth of 10~GHz so that
the Johnson-Nyquist current fluctuations in the highly resistive
contacts are of the order of 10~nA. This value is comparable with
the difference between $I_R$ and $I_L$ hence sets an lower bound for
the pulse widths. In order to improve the performance one should
either improve the CA as discussed previously, use stronger
excitations in order to improve the signal to noise ratio, or use
longer light pulses to decrease the noise bandwidth.

As stated above, in order to increase the repetition rate higher
conductance barriers are favorable so that the current transients
decay quickly after each pulse. Maximal CA is achieved for
comparable resistance of the semiconductor layer and of the contacts
but for too conductive barriers the optimal contact width may become
impractically small. In the upper panel the time-integrated CA is
about $6$$\%$ whereas the peak to peak ratio is about $9$$\%$. In
the lower panel both values are around $11$$\%$ which is probably
due to shorter dwelling time in the channel compared with the pulse
width (closer to the steady state solution). We have verified that
the time independent solution is recovered for pulses whose duration
is much longer than $\tau_{sp}$.

\subsection*{Ways to increase the current asymmetry}

CA may be improved by increasing the spin diffusion length $L_{sc}$
or by increasing the efficiency of the optically excited spin
polarization $\rho$. The latter may be increased from the bulk value
by lifting the degeneracy between heavy and light hole bands by
strain or by quantum well confinement.
\cite{Optical_Orientation,Marie_PRB00,Ohno_PRL99}.

Another alternative is to use barriers of higher conductance with
smaller contact widths yet still within lithography resolution
abilities (e.g. $\alpha$$\simeq$20 and
$w_{opt}$$\simeq$$0.1$$L_{sc}$$\simeq$$100$nm). We note, however,
that in the realm of current experiments,
\cite{Hanbicki_APL03,parkin} acquiring relatively high selectivity
occurs when the ferromagnet/semiconductor interface is abrupt,
\cite{Jonker_IEEE03} a situation which is achieved with barriers
whose $\alpha$ parameter is of the order of unity if non-degenerate
semiconductor channels are used. This is not the case in MOSFET
devices where the source and drain contacts are alloyed into the
semiconductor resulting with a low resistivity but at the expense of
a rough interface. \cite{Sze} A possible way to overcome the
limitation of low $\alpha$ while still acquiring high finesse is to
reduce the Schottky barrier thickness. This could be realized by
replacing the silicon dopant at the highly doped interface by
tellurium, tin or other dopants. \cite{Korshunov_compensation} For
these cases it was measured that self compensation occurred at
higher doping levels and that the free carrier concentration was
increased by a factor of four. This should halve the thickness of
barriers which are presently being used in spin injection
experiments, leading to an exponential increase in the contact
conductance.

Improvement of the spin selectivity
($|\beta|$$\rightarrow$$\alpha$), as demonstrated by replacing the
Schottky barrier with an insulating layer \cite{parkin} of course
helps but it was achieved at the expense of a lower $\alpha$
parameter since the conductance of the barrier was very low. This
could be improved if a thinner insulator layer is used.

\section{Conclusions} \label{sec:conclusions}

We have investigated the time-dependent response of spin diffusion
to light stimulation in a realistic lateral structure, which leads
to a simple spintronics-based scheme for the electrical measurement
of circular light polarization. The simulations were performed using
experimentally verified properties of a lateral Fe/GaAs system with
thin Schottky barriers. The results imply that spin accumulation in
short channels could be tracked in time with relatively high time
resolution when the Schottky barriers and the geometry are designed
properly. Our analysis provides an aid in choice of system
parameters which optimize the detection efficiency. Room temperature
detection is possible in short channels or in nanostructures where
the current optical detection techniques would be limited by
wavelength resolution.

This work is supported by NSF under Grant No. DMR-0325599.

%%%%%%%%%%%%%%%%%%%%%%%%%%%%%%%%%%%%%%%%%%%%%%%%%%%%%%%%%%%%%%%%%%%%%%%%%%%
\end{document}